# Zero-bias Anomaly of Tunneling into the Edge of a 2D Electron System.

L. Shekhtman and L. I. Glazman

*Theoretical Physics Institute, School of Physics and Astronomy, University of Minnesota,*

*116 Church Str. SE, Minneapolis, MN 55455*

(June 29, 1995 ; 14:49)

## Abstract

We investigate the electron tunneling into the edge of a clean weakly interacting two-dimensional electron gas. It is shown that the corresponding differential conductance $G(V)$ has a cusp at zero bias, and is characterized by a universal slope $|dG/dV|$ at $V = 0$. This singularity originates from the electron scattering on the Friedel oscillation caused by the boundary of the system.

It is well known that the electron-electron interaction in one dimension leads to a power-law singularity in the tunneling conductance at zero bias (see, e.g., [1]). When the interaction is weak, this anomaly may be attributed [2] to a singular backscattering of the incident electrons on the Hartree-Fock potential associated with the Friedel oscillation formed near the tunnel barrier. The Friedel oscillation affects the electrons on the Fermi surface almost like a periodic potential affects electrons with the wave vector on the boundary of the Brillouin zone. The tunneling density of states does not acquire a gap of finite width because unlike a strictly periodic potential, the oscillation decays with the distance from the barrier. However, this decay in 1D is sufficiently slow (inversely proportional to the distance) to lead to a power-law singularity in the density of states at the Fermi level. It has been shown [3] that this power-law singularity survives also in a multichannel case – for tunneling into the edge of a long wire of a finite width. However, the corresponding exponent decreases with an increase in the number of the transverse modes. The question is, whether any reminiscence of this zero-bias anomaly persists in the limit of an infinite number of modes, i.e., in two dimensions.

We show in this paper that the differential conductance for tunneling into the edge of a clean, interacting two-dimensional electron gas (2DEG) is singular at zero bias. The backscattering from the Friedel oscillation in this case does not renormalize the zero-bias conductance to zero, but still leads to a non-analytic behavior of $G(V)$ at small applied voltages. The corresponding cusp in $G(V)$ is characterized by a finite slope $|dG/dV|$ at $V = 0$, which is universal, i.e., does not depend on the interaction strength.

To clarify the origin of the cusp in the differential conductance, we investigate first tunneling into the edge of 2DEG with a short-range electron-electron interaction. In this case, the lowest order perturbation theory in the interaction potential is applicable. Our calculation of conductance in a weakly interacting system is based on the Landauer [4] approach. Following Ref. [2] we will focus on the effect of electron-electron interactions on the transmission coefficient for tunneling through a one-dimensional barrier separating two semi-planes. We assume that the transmission coefficient of the barrier is small, $|t_0|^2 \ll 1$, and that the barrier is homogeneous along the $y$-direction [5]. In the absence of interaction, the current $I$ per unit length of the barrier may be obtained from a straightforward generalization of the Landauer formula:

$$I = \frac{2e}{\hbar} \int_{-\infty}^{\infty} \frac{d\epsilon}{2\pi} [f_l(\epsilon - eV) - f_r(\epsilon)] \int_{-\infty}^{\infty} \frac{dk_y}{2\pi} |t_0(k_y, \epsilon)|^2, \qquad (1)$$

where $f_l(\epsilon - eV)$ and $f_r(\epsilon)$ are the Fermi distribution functions in the left and right semi-planes respectively, and $\epsilon$ is the kinetic energy of an electron. At small bias, Eq. (1) yields

$$I(V) = \frac{e^2}{2\pi\hbar} k_F T_0 V, \qquad (2)$$

where $k_F$ is the Fermi wave vector in the 2DEG, and $T_0 = \langle |t_0|^2 \rangle$ is the bare transmission coefficient at the Fermi surface averaged over the directions of momenta of the incoming electrons. The bare transmission coefficient may depend on energy, which gives rise to the well-known field effect in tunneling. Therefore, the r.h.s. of Eq. (2) can be viewed as a linear term in expansion of the $I(V)$ function, the next term being proportional to $V^3$. Unlike this one-particle field effect, the electron-electron interaction leads to a stronger and non-analytical correction to the current (2).

The existence of the barrier breaks the translation invariance in the $x$ direction, leading to the Friedel oscillation of the electron density. Due to this oscillation, the electron-electron interaction produces an additional potential, $\hat{V}_{\text{eff}}(x;\mathbf{k})$, which enhances the backscattering of electrons. The corresponding correction to the transmission amplitude, $\delta t(\mathbf{k})$, can be found in the Born approximation. To do this, we notice that the component of the electron momentum parallel to the barrier is conserved, and, correspondingly, the electron wave function has the form $\Psi_{\mathbf{k}}(x,y) = e^{ik_y y}\psi_{\mathbf{k}}(x)$. In the absence of the electron-electron interaction, the wave function $\psi_{\mathbf{k}}(x) \equiv \phi_{k_x}(x)$ is the scattering state formed by the bare barrier potential. The correction found in the first order in the interaction potential is

$$\delta\psi_{\mathbf{k}}(x) = \int_{-\infty}^{\infty} g_{k_x}(x,x')\hat{V}_{\text{eff}}(x';\mathbf{k})\phi_{k_x}(x')dx', \tag{3}$$

where $g_{k_x}(x,x')$ is the Green function of the non-interacting single-electron Hamliltonian in the presence of the barrier. It has the following asymptotic form at $x' < x$, $x \to \infty$:

$$g_{k_x}(x,x') = \begin{cases} \frac{m}{i\hbar^2 k_x}\left[e^{ik_x(x-x')} + r_0 e^{-ik_x(x+x')}\right], & x' > 0, \\ \frac{m}{i\hbar^2 k_x}t_0 e^{ik_x(x-x')}, & x' < 0, \end{cases} \tag{4}$$

where $r_0$ is the reflection amplitude. For the wave incoming from $x < 0$, Eq. (3) gives the correction of the form

$$\delta\psi_{\mathbf{k}}(x) \simeq \frac{1}{\sqrt{2\pi}}\delta t(\mathbf{k})e^{ik_x x}, \quad x \to +\infty, \tag{5}$$

which defines $\delta t(\mathbf{k})$. The non-analytic contribution to $\delta t(\mathbf{k})$ is determined by the Friedel oscillation present in the asymptotic behavior of the effective potential at large $x$. This oscillation is caused by the reflection off the barrier of all the electrons forming the Fermi sea, and is characteristic for both Hartree- and exchange-type terms in $\hat{V}_{\text{eff}}(x;\mathbf{k})$:

$$\hat{V}_{\text{eff}}(x;\mathbf{k}) = V_H(x) + \hat{V}_{\text{ex}}(x;\mathbf{k}). \tag{6}$$

The Hartree term is local,

$$V_H(x) = \int U_{ee}(\mathbf{r}-\mathbf{r}_1)n(\mathbf{r}_1,\mathbf{r}_1)d\mathbf{r}_1, \tag{7}$$

and the exchange term is given by the following integral relation:

$$\hat{V}_{\text{ex}}(x;\mathbf{k})\phi_{k_x}(x) = \int U_{ee}(\mathbf{r}-\mathbf{r}_1)n(\mathbf{r},\mathbf{r}_1)\phi_{k_x}(x_1)e^{ik_y(y-y_1)}d\mathbf{r}_1. \tag{8}$$

Here $U_{ee}(\mathbf{r}-\mathbf{r}_1)$ is the electron-electron interaction potential, and the electron density matrix is given by

$$n(\mathbf{r},\mathbf{r}_1) = \int \frac{d\mathbf{q}}{2\pi^2}n_F(\mathbf{q})\phi_{q_x}^*(x_1)\phi_{q_x}(x)e^{iq_y(y-y_1)}, \tag{9}$$

where $n_F(\mathbf{q}) = \theta(k_F - q)$ is the zero-temperature Fermi distribution function, and $k_F$ is the Fermi wavevector. The discontinuity in the electron momentum occupation number $n_F(\mathbf{q})$ results in the Friedel oscillation of the density $n(\mathbf{r}, \mathbf{r}_1)$, and in the effective potential $\hat{V}_{\text{eff}}(x; \mathbf{k})$. The latter becomes local,

$$V_H(x) \simeq \tilde{U}_{ee}(2k_F) \frac{8k_F^2}{\pi(2k_F|x|)^{3/2}} \sin(2k_F|x| + \frac{3}{4}\pi), \tag{10}$$

$$V_{\text{ex}}(x; \mathbf{k}) \simeq \tilde{U}_{ee}(0) \frac{8k_F^2}{\pi(2k_F|x|)^{3/2}} \sin(2k_F|x| + \frac{3}{4}\pi), \tag{11}$$

for distances $x$ from the barrier exceeding both the Fermi wavelength $\lambda_F$ and the range $d$ of the potential $U_{ee}(\mathbf{r})$. In derivation of (10) and (11) we assumed that the transmission coefficient is small, $|t_0|^2 \ll 1$, and the components of $\mathbf{k}$ satisfy the conditions: $|k_x - k_F| \ll k_F$ and $k_y \ll k_F$. The latter two conditions allowed us to express the amplitudes of oscillation in (10), (11) in terms of Fourier harmonics $\tilde{U}_{ee}(2k_F)$, $\tilde{U}_{ee}(0)$ of the interaction potential $U_{ee}(\mathbf{r})$. Substituting now $U_{\text{eff}}(x; \mathbf{k})$ into (3), we can find the correction $\delta t(k_x, k_y)$. The oscillation in the potential (10), (11) leads to the non-analytical at $k_x = k_F$ part of $\delta t(k_x, k_y)$. The transmission coefficient $|t(\mathbf{k})|^2$ for electrons with energies close to the Fermi level to the first order in the strength of the electron-electron interaction is given by

$$|t(\mathbf{k})|^2 = |t_0(\mathbf{k})|^2 \left\{ 1 - \alpha + \frac{m}{4\pi^2 \hbar^2} \left[ \tilde{U}_{ee}(0) - \tilde{U}_{ee}(2k_F) \right] \sqrt{\frac{k_x - k_F}{k_F}} \theta(k_x - k_F) \right\}. \tag{12}$$

Here we presented explicitly only the non-analytical part of the dependence of the transmission coefficient on the incident wavevector. The bare transmission coefficient, $|t_0(\mathbf{k})|^2$, and the part of the correction $\alpha \sim m[\tilde{U}_{ee}(0) - \tilde{U}_{ee}(2k_F)]/\hbar^2$ are regular at $k = k_F$ functions.

In order to calculate the corresponding contribution to the conductance, we have to substitute (12) instead of $|t_0|^2$ in Eq. (1). The calculation then amounts to integrating the transmission coefficient over the energies that are larger than $\epsilon_F$ but smaller than $\epsilon_F + eV$, where $\epsilon_F$ is the Fermi energy and $eV$ is the applied bias. It follows from (12) that only the electrons incoming with the momentum $k_x$ larger than the Fermi wavevector contribute to the non-analytic correction. At $eV \ll \epsilon_F$ these are electrons moving in the small range of incident angles almost perpendicular to the barrier, and the corresponding bare transmission coefficient is $|t_0(k_y = 0, \epsilon_F)|^2$. Averaging Eq. (12) over all the incident angles, we obtain the expression for the differential conductance $G(V) \equiv dI/dV$ per unit length of the barrier at small biases:

$$G(V) = G_0 \left( 1 - \alpha + \gamma \frac{m}{\hbar^2} \left[ \tilde{U}_{ee}(0) - \tilde{U}_{ee}(2k_F) \right] \frac{|eV|}{\epsilon_F} \right). \tag{13}$$

The non-analytical part in (13) provides a cusp in the differential conductance at zero bias. The numerical coefficient $\gamma \approx |t_0(k_y = 0, \epsilon_F)|^2/(2\pi)^2 T_0$ accounts for the difference between the average transmission coefficient $T_0$ that determines $G_0$, and the relevant for the anomaly transmission coefficient $|t_0(k_y = 0, \epsilon_F)|^2$. The value of $\gamma$ depends on the detailed shape of the barrier, but typically $\gamma \simeq 1$.

We are allowed to treat the potential energy of electron-electron interaction as a small perturbation only under the conditions

$$\frac{m\tilde{U}_{ee}(2k_F)}{\hbar^2} \ll 1, \tag{14}$$

$$\frac{m\tilde{U}_{ee}(0)}{\hbar^2} \ll 1. \tag{15}$$

Hence, the "bare" interaction must be weak *and short-ranged*. This latter condition can be implemented in experiments, if a metallic gate exists very close to the 2D electron gas. The interaction between electrons is described by the Coulomb potential only at distances $r$ smaller than the separation $d$ between the 2D electron gas and the gate. At $r \gg d$ the potential $U_{ee}(r)$ has a dipolar asymptotic behavior, and the condition (15) is met if $d$ is much smaller than the effective Bohr radius, $a_B = \varepsilon\hbar^2/me^2$, where $\varepsilon$ is the dielectric constant of the semiconductor. Eq. (13) leads to the following estimate of the strength of the cusp:

$$\frac{dG}{dV} \simeq \left(\frac{d}{a_B}\right) G_0 \text{sign}(eV). \tag{16}$$

In the most interesting case of a GaAs heterostructure, the thickness of a spacer separating the 2D electron gas from the gate normally exceeds $a_B$. We will show that at $d \gg a_B$ the proportional to the interaction strength factor $d/a_B$ in the estimate (16) is replaced by an interaction-independent universal constant.

In the limit $d \gg a_B$, for a pure long-range Coulomb interaction, the requirement (14) is equivalent to the condition that the plasma parameter $e^2/\varepsilon\hbar v_F$ is small. This condition is satisfied in a sufficiently dense interacting electron gas, and the above approach correctly gives the leading order Hartree-type contribution to the differential conductance. However, the condition (15) does not hold at any density since $\tilde{U}_{ee}(k \to 0)$ diverges. Therefore the slope of the cusp in $G(V)$ might be not small, and the exchange contribution to the Eq. (13) must be revised. Specifically, we have to go beyond the perturbative single-electron picture presented above and to consider the many-body effects leading to screening of the long-range part of the Coulomb interaction.

To incorporate the screening effects, we use the standard [6] relation of the tunneling current $I$ with the product of the densities of states, which is valid at small $|t_0|^2$. This relation can be cast in the form (1), if one replaces the bare transmission coefficient $|t_0|^2$ by the renormalized value:

$$T(k_y, \epsilon) = |t_0(k_y, \epsilon_F)|^2 \frac{\pi^2 \hbar^4}{4m^2 K_x^2} \rho_l(k_y, \epsilon) \rho_r(k_y, \epsilon). \tag{17}$$

Here

$$K_x = k_x(k_y, \epsilon) \equiv \sqrt{(2m\epsilon/\hbar^2) - k_y^2} \tag{18}$$

is the $x$-component of the wavevector at given total energy $\epsilon$ and fixed value of $k_y$. The densities of electron states $\rho_{r(l)}(k_y, \epsilon)$ are related [7] to the corresponding single-electron Green functions $G_{r(l)}(x, x'; k_y; \epsilon)$ of the isolated right and left semiplanes respectively,

$$\rho_{r(l)}(k_y, \epsilon) = \text{Im} \frac{\partial^2}{\partial x \partial x'} \left(-\frac{1}{\pi} G_{r(l)}(x, x'; k_y; \epsilon)\right)\bigg|_{x=x'=0}. \tag{19}$$

The latter relation for the densities of states accounts for the fact that tunneling occurs into the *edge* of 2DEG, and the electron wavefunctions should vanish at the edge of an isolated semiplane,

$$G_{r(l)}(x, x'; k_y; \epsilon)\big|_{x=x'=0} = 0. \tag{20}$$

As we have shown above, the zero-bias anomaly in the differential conductance originates from the scattering of the incoming electrons on the Friedel oscillation. Although this oscillation is induced by the barrier, only its *tail* at large distances ($\sim \hbar^2 k_F/meV$) from the barrier is responsible for the anomaly. In the many-body formulation the potential of Friedel oscillation is represented by a specific coordinate dependence of the electron self-energy $\Sigma$. The latter contributes to the Green function,

$$\delta G(x, x'; k_y; \epsilon) = \int G^{(0)}(x, x_1; k_y; \epsilon) \Sigma(x_1, x_2; k_y; \epsilon) G^{(0)}(x_2, x'; k_y; \epsilon) dx_1 dx_2, \tag{21}$$

and, therefore, to the density of states. It is essential, that the self-energy in Eq. (21) depends not only on the coordinate difference $x_1 - x_2$, but also on the distance from the barrier, i.e., on the *sum* of the coordinates $x_1 + x_2$. Our aim is to investigate the induced by the barrier contribution to $\Sigma$, to show that it has the Friedel oscillation form, and to calculate the corresponding correction to the tunneling density of states. The outlined program based on the calculation of $\Sigma$, enables us to generalize Eq. (11) so that the long-range nature of the bare Coulomb interaction between the electrons is accounted for.

As may be anticipated from Eq. (11), the lowest order in electron charge $e$ exchange contribution to $\Sigma$ is proportional to $\tilde{U}_{ee}(k \to 0)$. For the long-range Coulomb interaction, this Fourier harmonic diverges. The standard way to deal with this difficulty is to sum the most singular at $k \to 0$ diagrams in *each* order of the self-energy expansion in $\tilde{U}_{ee}(\mathbf{k})$. In the translationally invariant system it is easy to identify these most singular diagrams, since for such a system the dielectric function of the electron gas $\kappa(\mathbf{r_1} - \mathbf{r_2}; \epsilon)$ is diagonal in the plane-wave representation (see, e.g., Ref. [6]). It is well-known [8] that the described summation is equivalent to the replacement of $\tilde{U}_{ee}(\mathbf{k})$ in the formula for the leading-order contribution to the exchange part of the self energy by the effective *screened* interaction. The latter is given by $\tilde{U}_{ee}(\mathbf{k})/\tilde{\kappa}(\mathbf{k}; \epsilon)$, where $\tilde{\kappa}(\mathbf{k}; \epsilon)$ is a Fourier transform of $\kappa(\mathbf{r_1} - \mathbf{r_2}; \epsilon)$. In our problem, however, the translational invariance is destroyed by the barrier. The dielectric function $\kappa(\mathbf{r_1}, \mathbf{r_2}; \epsilon)$ depends not only on $\mathbf{r_1} - \mathbf{r_2}$, but also on the distance from the barrier, i.e., on $x_1 + x_2$. Thus, the problem of diagonalization of the dielectric function becomes non-trivial, and the screened electron-electron interaction potential in the vicinity of the barrier can not be found in a closed form.

Our task is simplified by the fact that we do not need to know this expression for the screened electron-electron interaction in the vicinity of the barrier. As we noted already, the zero-bias anomaly arises due to the scattering of the incident electrons by the Friedel oscillation *far from the barrier*, at the distances $\sim \hbar^2 k_F/meV$. Thus, we are interested in the asymptotic form of the self energy $\Sigma(x_1, x_2; k_y; \epsilon)$ at $x_1, x_2 \gg a_B, \lambda_F$. There the dielectric function loses all the information about the barrier and becomes a function of $|x_1 - x_2|$ only. Thus, while calculating the singular contribution to the electron self-energy, we may use the formulas for the dielectric function and, hence, for the screened electron-electron interaction in the translationally invariant case. In dense electron liquid ($e^2/\varepsilon \hbar v_F \ll 1$) this screened electron-electron interaction is well described by the expression found in RPA:

$$U_{RPA}(x_1 - x_2, k_y; \epsilon) = \int dq_x e^{iq_x(x_1-x_2)} \frac{\tilde{U}_{ee}(q_x, k_y)}{\tilde{\kappa}(q_x, k_y; \epsilon)}, \qquad (22)$$

where $\tilde{\kappa}(q_x, k_y; \epsilon)$ is a two-dimensional dielectric function of the translation invariant electron gas, see, e. g., [9]. Below we concentrate on the leading order exchange-type term

$$\Sigma(x_1, x_2; k_y; \epsilon)|_{x_1, x_2 \gg a_B, \lambda_F} = -i \int dq_y d\epsilon' U_{\text{RPA}}(x_1 - x_2; k_y - q_y; \epsilon - \epsilon') G^{(0)}(x_1, x_2; q_y; \epsilon') \qquad (23)$$

since, as we already noted, in the considered dense electron liquid $(e^2/\varepsilon \hbar v_F \ll 1)$ the Hartree-type contribution is correctly accounted for by the formula (13).

It is easy to identify now the origin of the zero-bias anomaly. In Eq. (23) all the relevant for singularity information about the presence of the barrier is delivered by the Green function. Specifically, the oscillating part of the dependence of $\Sigma$ on its coordinates originates in the oscillatory dependence of the Green function on the *sum* of the coordinates $x_1 + x_2$. It is easy to see, that far from the barrier $G^{(0)}(x_1, x_2; q_y; \epsilon)$ may be decomposed as

$$G^{(0)}(x_1, x_2; q_y; \epsilon)\Big|_{x_1, x_2 \gg \lambda_F} \approx G^+(x_1 + x_2; q_y; \epsilon) + G^-(x_1 - x_2; q_y; \epsilon),$$

and the singular contribution to the self-energy has the form:

$$\Sigma^{sing}(x_1, x_2; k_y; \epsilon) = -i \int dq_y d\epsilon' U_{\text{RPA}}(x_1 - x_2; k_y - q_y; \epsilon - \epsilon') G^+(x_1 + x_2; q_y; \epsilon'). \qquad (24)$$

Straightforward calculations based on Eqs. (19), (21) and (24) lead to the following expression for the anomalous contribution to the tunneling density of states:

$$\delta\rho(k_y; \epsilon) = \text{Im}\left(\frac{4m^2 k_F^2}{\pi^2 \hbar^4} \int_d^\infty dx\, e^{2iK_x x} V_{\text{ex}}^{\text{eff}}(x; k_y; \epsilon)\right), \qquad (25)$$

where

$$V_{\text{ex}}^{\text{eff}}(x, k_y; \epsilon) = \tilde{U}_{RPA}\left(K_x - k_F, k_y; \epsilon - \epsilon_F\right) \sqrt{\frac{2}{\pi}} \frac{e^{-2ik_F x}}{(2k_F x)^{3/2}}, \qquad (26)$$

wave vector $K_x \equiv k_x(k_y, \epsilon)$ is given by (18), and $\tilde{U}_{RPA}(\mathbf{q}; \epsilon) = \tilde{U}_{ee}(\mathbf{q})/\kappa(\mathbf{q}; \epsilon)$.

If the applied bias is small $[eV < (e^2/\varepsilon \hbar v_F)\epsilon_F]$, we may neglect the energy and wavevector dependence of the screened interaction in Eqs. (25) and (26), and take $\tilde{U}_{RPA}(\mathbf{q}; \epsilon) \approx \tilde{U}_{RPA}(0; 0)$. Now we note that $\tilde{U}_{RPA}(0; 0) = \hbar^2/m$ is independent of the interaction strength. In complete analogy with the problem of tunneling into a disordered interacting 2DEG [10], this fact leads to universality of the correction to the electron density of states. Substituting Eq. (26) into (25) we find for the correction to the density of states at energies close to $\epsilon_F$:

$$\delta\rho(k_y; \epsilon) = \frac{1}{\pi^2}\left(\frac{2m}{\hbar^2}\right)^{\frac{3}{2}} \left(\epsilon - \frac{\hbar^2 k_y^2}{2m} - \epsilon_F\right)^{1/2} \theta\left(\epsilon - \frac{\hbar^2 k_y^2}{2m} - \epsilon_F\right). \qquad (27)$$

Thus, the anomalous exchange-type contribution of electron-electron interaction to the transmission coefficient is given by

$$\delta T(k_y; \epsilon) \propto |t_0(k_y, \epsilon_F)|^2 \frac{\delta\rho(k_y; \epsilon)}{\rho_0(k_y; \epsilon)} = \frac{|t_0(k_y, \epsilon_F)|^2}{\pi} \left(\epsilon - \frac{\hbar^2 k_y^2}{2m} - \epsilon_F\right)^{1/2} \theta\left(\epsilon - \frac{\hbar^2 k_y^2}{2m} - \epsilon_F\right), \tag{28}$$

where $\rho_0(k_y; \epsilon) = 2mk_x(k_y; \epsilon)/\pi\hbar^2$ is the density of states (19) for tunneling into an edge of non-interacting electron system. Substituting (28) in (1) we find the expression for the differential conductance to the first order on the screened electron-electron interaction:

$$G(V) = G_0\left(1 - \tilde{\alpha} + \gamma\frac{|eV|}{\epsilon_F}\right), \tag{29}$$

where $\tilde{\alpha} \sim e^2/\varepsilon\hbar v_F$ is small and $\gamma \simeq 1$. We neglected in Eq. (29) the Hartree-type contribution [see Eq. (13)], since its ratio to the exchange-type contribution is of the order of $\tilde{U}_{RPA}(2k_F)/\tilde{U}_{RPA}(0) \approx e^2/\varepsilon\hbar v_F$, and therefore small. Clearly, at a larger bias, $eV \gtrsim (e^2/\varepsilon\hbar v_F)\epsilon_F$, one can not approximate $\tilde{U}_{RPA}(\mathbf{q}; \epsilon)$ in (26) by $\tilde{U}_{RPA}(0; 0)$, since the screening is not effective at $|\mathbf{q}| \gtrsim 1/a_B$. At these biases the exchange contribution to differential conductance becomes comparable with the Hartree contribution, the slope of the curve $G(V)$ decreases, and the interaction-induced correction to the conductance eventually vanishes. The qualitative dependence of the tunneling differential conductance on the applied bias is shown in Fig. 1.

For simplicity we considered above the case of the uniform barrier, where the parallel to the barrier component of the tunneling electron momentum is conserved. It is clear, however, that our result does not depend on this barrier uniformity. Indeed, the zero-bias anomaly is not sensitive to the changes in the bare transmittion coefficient, and originates exclusively from the singularity in the electron density of states $\rho(k_y; \epsilon)$ at the Fermi surface. It is easy to check, that even if the barrier is not uniform, the tunneling current to leading order in electron-electron interaction is still given by the formula

$$I = \sum_{i=r,l} \frac{2e}{\hbar} \int_{-\infty}^{\infty} \frac{d\epsilon}{2\pi} [f_l(\epsilon - eV) - f_r(\epsilon)] \int_{-\infty}^{\infty} \frac{dk_y^i}{2\pi} F(k_y^i; \epsilon) \left(1 + \frac{\delta\rho^i(k_y^i; \epsilon)}{\rho_0^i(k_y^i; \epsilon)}\right), \tag{30}$$

where the superscript $i = r, l$ refers to the left or right semiplane respectively. Clearly, the singular behavior of the density of states $\delta\rho^i(k_y^i; \epsilon)$ at the Fermi surface (at $k_y^i = 0$) does not depend on details of tunneling. This behavior is adequately accounted for by Eq. (27). All the information about the shape of the barrier, i.e., about the expression for the bare transmission coefficient $|t_0(\mathbf{k}^r, \mathbf{k}^l; \epsilon)|^2$, is absorbed in the factor $F(k_y^i; \epsilon)$. It is obvious, that $F(k_y^i; \epsilon)$ is a regular function of energy and momentum at $\epsilon = \epsilon_F$ and $k_y^i = 0$. This directly leads to Eq. (29).

In conclusion, we demonstrated that a weak interaction in the clean two-dimensional electron gas leads to a singular contribution to tunneling into the edge of electron system. This singularity is characterized by a non-analytical behavior of the corresponding density of states near the Fermi level, and leads to a characteristic universal cusp in the differential conductance at zero bias.

The authors are grateful to I. L. Aleiner for critical reading of the manuscript and useful comments. This work was supported by NSF Grant DMR-9423244.

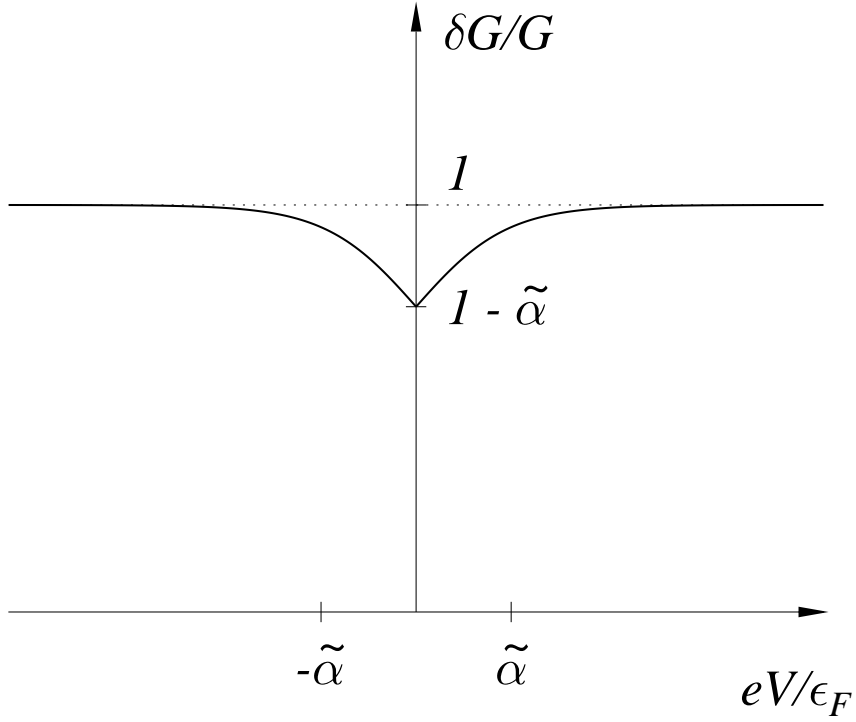

FIG. 1. The schematic dependence of the differential conductance $G(V) = dI/dV$ on the applied bias $V$ for tunneling into the edge of interacting two-dimensional electron gas. The amplitude of the interaction-induced correction is proportional to the interaction strength and small, $\tilde{\alpha} \simeq e^2/\varepsilon\hbar v_F$, and the correction vanishes at larger bias. However the cusp in $G(V)$ at $V=0$ does not depend on the interaction strength. A smooth variation of $G(V)$ due to the field effect on the shape of the tunneling barrier is not shown.